\documentclass[apjl,iop]{emulateapj}
\usepackage{times,mathptmx,color}

\shorttitle{Outward migration of Jupiter and Saturn in radiative disks}
\shortauthors{Pierens et al.}

\begin{document}


\title{Outward migration of Jupiter and Saturn in 3:2 or 2:1 resonance in radiative disks: 
implications for the Grand Tack and Nice models}


\author{Arnaud Pierens \altaffilmark{1,2}, Sean ~N. Raymond\altaffilmark{1,2}, David Nesvorny\altaffilmark{3}, Alessandro  Morbidelli\altaffilmark{4}}
\altaffiltext{1}{Univ.  Bordeaux, Laboratoire d'Astrophysique de Bordeaux,
    UMR 5804, F-33270, Floirac, France }
    
\altaffiltext{2}{CNRS, Laboratoire d'Astrophysique de Bordeaux, UMR 5804, F-33270, Floirac, France}

\altaffiltext{3}{Department of Space Studies, Southwest Research Institute, 1050 Walnut Street Suite 300, Boulder, 
CO 80302, USA}

\altaffiltext{4}{ University of Nice-Sophia Antipolis, CNRS, Observatoire de la c\^ote d'Azur, 
Laboratoire Lagrange, BP4229, 06304 NICE Cedex 4, France}



\begin{abstract}

Embedded in the gaseous protoplanetary disk, Jupiter and Saturn naturally become trapped in 3:2 resonance and migrate outward.  This serves as the basis of the Grand Tack model.   However, previous hydrodynamical simulations 
were restricted to isothermal disks, with moderate aspect ratio and viscosity.  Here we simulate the orbital evolution of the gas giants in disks with viscous heating and radiative cooling.  We find that Jupiter and Saturn migrate outward in 3:2 resonance in modest-mass ($M_{disk} \approx M_{MMSN}$, where MMSN is the "minimum-mass solar nebula") disks with viscous stress parameter $\alpha$ between $10^{-3}$ and $10^{-2} $.   In disks with relatively low-mass ($M_{disk} \lesssim M_{MMSN}$) ,  Jupiter and Saturn get captured  in 2:1 resonance and can even 
migrate outward in low-viscosity  disks ($\alpha \le 10^{-4}$). Such disks have a very small aspect ratio
($h\sim 0.02-0.03$) that   favors outward migration after 
capture in 2:1 resonance, as confirmed by  isothermal runs which  resulted in a similar outcome for $h \sim 0.02$ 
and $\alpha \le 10^{-4}$.   We also performed N-body runs of the outer Solar System starting from the results of our 
hydrodynamical  simulations and including 2-3 ice giants.  After dispersal of the gaseous disk, a Nice model instability starting with Jupiter and Saturn in 2:1 resonance results in  good Solar Systems analogs.  We conclude that in a  cold Solar Nebula, the 2:1 resonance between Jupiter and Saturn can lead to outward migration of the system, and this may represent an alternative scenario for the evolution of the Solar System. 

\end{abstract}





\section{Introduction}

Two new models piece together a story of the evolution of the Solar System.  They are the Grand Tack and Nice models.

The Grand Tack model addresses the formation of the inner Solar System (Walsh et al 2011; see also Morbidelli et al 2012; Raymond et al 2013; O'Brien et al 2014).   It invokes an inward-then-outward migration of Jupiter and Saturn consistent with hydrodynamical simulations (Masset \& Snellgrove 2001; Morbidelli \& Crida 2007).  The change in the direction of migration, or "tack",  is set to  occur when Jupiter is at 1.5 AU,  so that  the sculpted inner  disk  of planetary embryos can reproduce the terrestrial planets, in particular the large Earth/Mars mass ratio (Wetherill 1991; Hansen 2009; Raymond et al 2009).  

The Nice model proposes that  after the gas disk dispersal, the outer Solar System underwent a delayed instability triggered by interactions between the giant planets and an outer disk of planetesimals.  In the original version of the model, the giant planets started on an arbitrary but more closely-packed orbital configuration. The instability was triggered by Jupiter and Saturn crossing their mutual 2:1 resonance (Tsiganis et al 2005). In more recent incarnations, the giant planets' orbital configuration was sculpted by an earlier phase of gas-driven migration (Morbidelli et al 2007),  with initial conditions that are consistent with the Grand Tack scenario.  Angular momentum exchange between the giant planets and the planetesimal disk can act to extract the giant planets from a resonant chain and trigger the instability (Levison et al 2011).  

The Nice model has been invoked to explain the Late Heavy Bombardment (Gomes et al. 2005). It can reproduce the current architecture of the giant planets (Nesvorny 2011; Nesvorny \& Morbidelli 2012), the orbital distribution of Jupiter's Trojan asteroids (Morbidelli et al. 2005; 
Nesvorny et al. 2013),  the irregular satellites of Jupiter (Nesvorny et 
al. 2014), Saturn, Uranus and Neptune (Nesvorny et al. 2007), and the structure of the Kuiper belt  in broad strokes (Levison et al. 2008, Batygin et al. 2012).

Jupiter and Saturn's gas-driven migration in 3:2 resonance has been confirmed by several studies (Masset \& Snellgrove 2001; Morbidelli \& Crida 2007; Pierens \& Nelson 2008; Pierens \& Raymond 2011).  Jupiter and Saturn naturally converge into 3:2 resonance and then migrate outward.  However, each of these studies used an  isothermal disk with a standard value for the aspect ratio $h\sim 0.05$.

In this paper we present the outcome of hydrodynamical simulations of the planets embedded in gaseous disks that are subject to viscous heating and radiative cooling.  We show that in  relatively low-mass ($M_{disk} \lesssim M_{MMSN}$, where MMSN is the minimum-mass solar nebula model of Hayashi 1981) and low-viscosity (viscous stress parameter $\alpha = 10^{-4}$ or $10^{-5}$; see Shakura \& Sunyaev 1973) disks, Jupiter and Saturn can migrate outward in 2:1 resonance.  This arises because for a purely viscously heated disk, a low-disk mass and/or a small viscosity give rise to a very small disk aspect ratio  $h\sim 0.02-0.03$, which tends to favor outward migration.  The advantage of having Jupiter and Saturn in 2:1 rather than in 3:2  resonance is that it can increase the success rate of the Nice model instability.  Nesvorny \& Morbidelli (2012) indeed showed that if Jupiter and Saturn start in a 2:1 resonance, 
there is a better chance that their  period ratio jumps to the correct value.

Our paper is structured as follows. In Sect. 2, we describe the hydrodynamical model and discuss the results 
of our simulations in Sect. 3,  and the effects of stellar heating on these results in Sect. 4.  Nice model simulations that 
study the evolution after disk dispersal are presented in Sect. 5.  Finally, we draw our conclusions in Sect. 6.

\begin{figure}
\includegraphics[width=0.8\columnwidth]{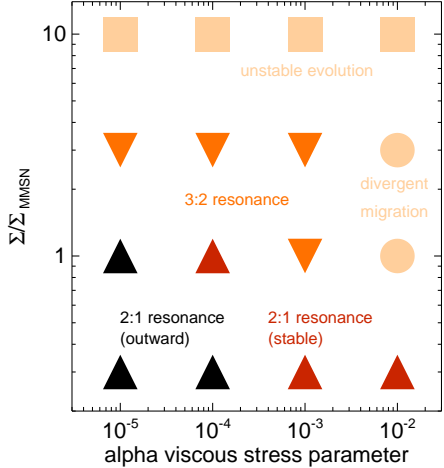}
\caption{ Results of the $16$ simulations ({\it symbols}) we performed as a function of the $\alpha$ viscous stress parameter 
and the  normalized initial disk surface density at $1$ AU.}
\label{fig:bilan}
\end{figure}

\begin{figure*} 
\centering
\includegraphics[width=0.33\textwidth]{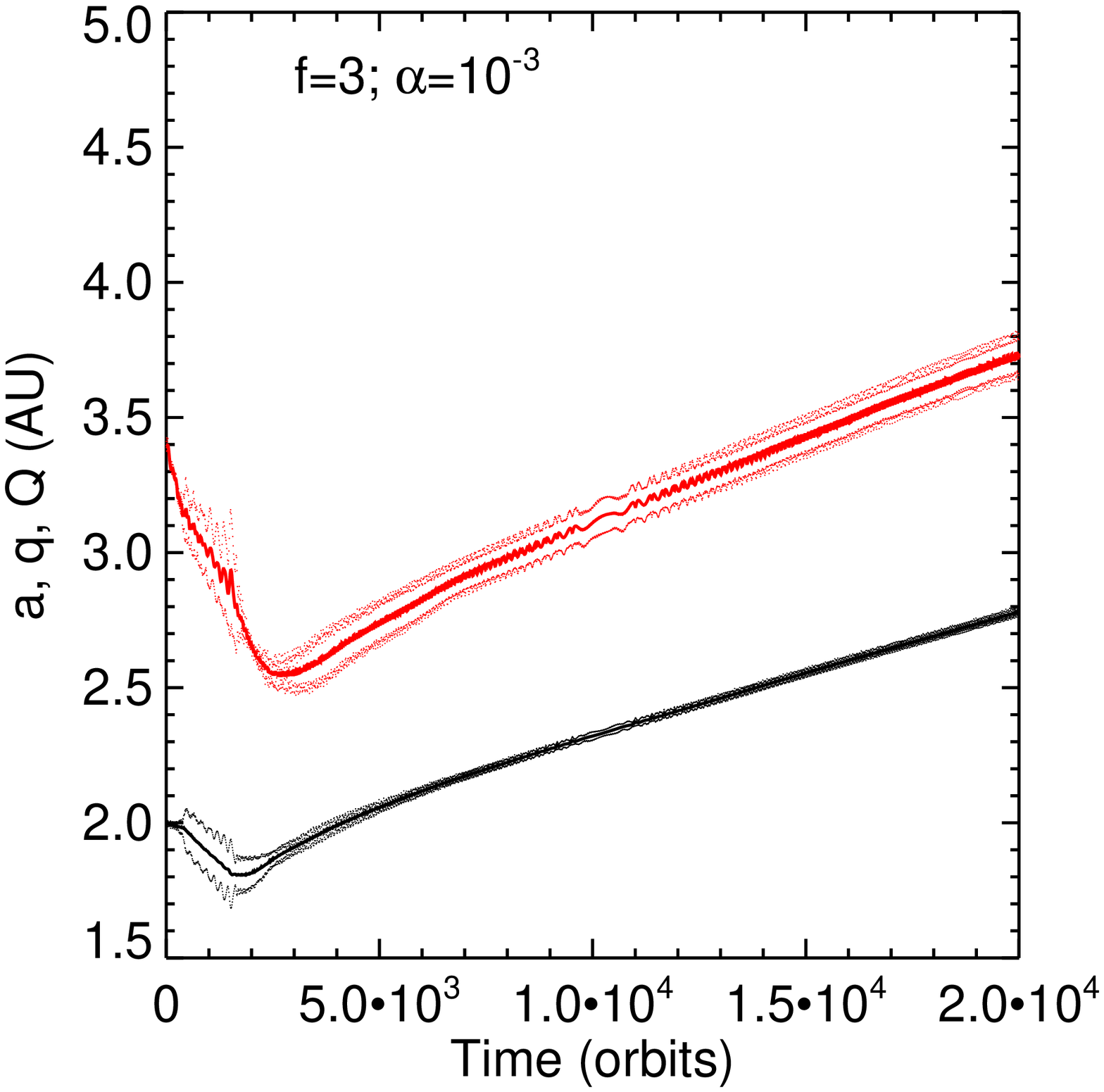}
\includegraphics[width=0.33\textwidth]{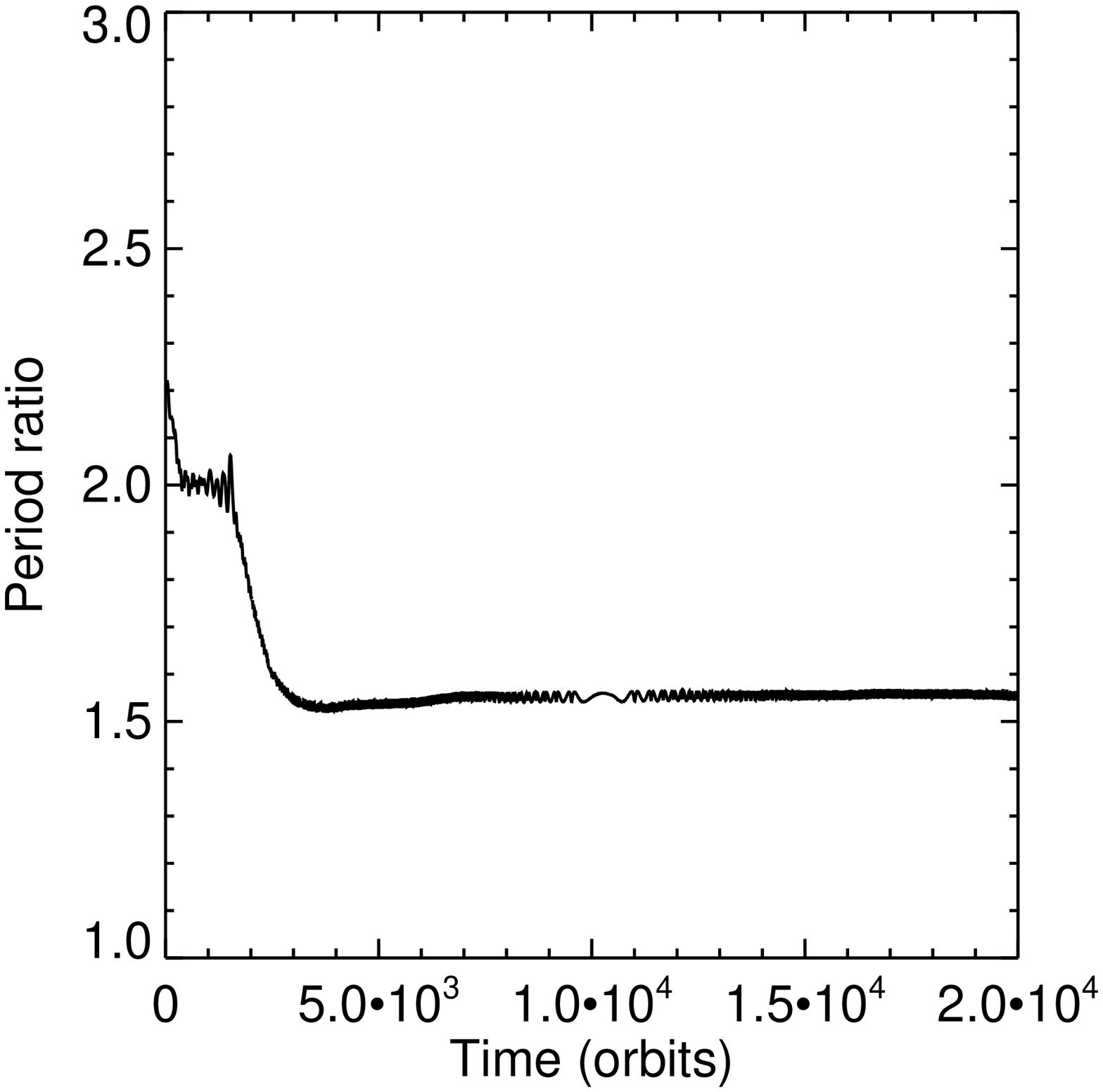}
\includegraphics[width=0.33\textwidth]{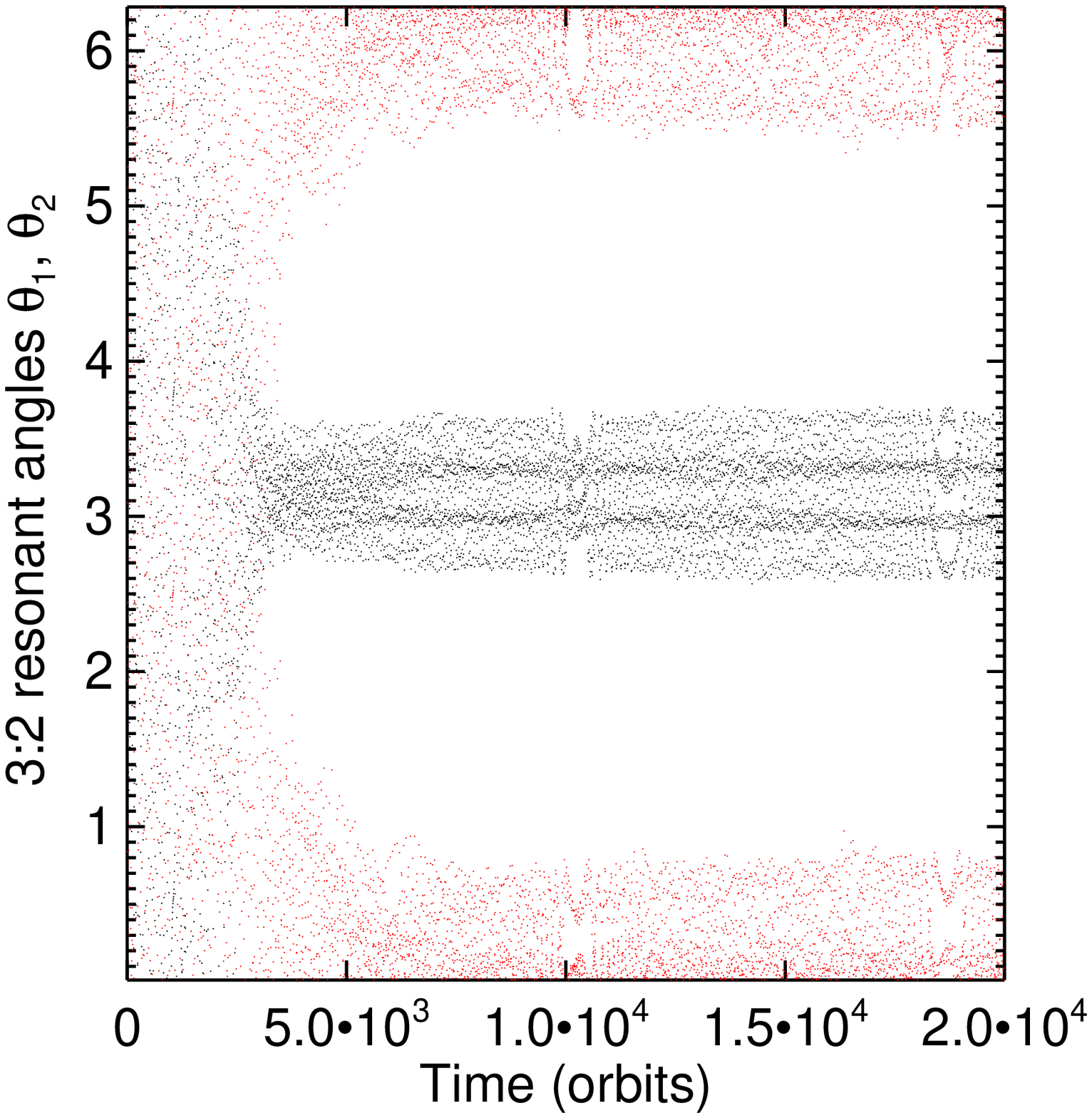}
\includegraphics[width=0.33\textwidth]{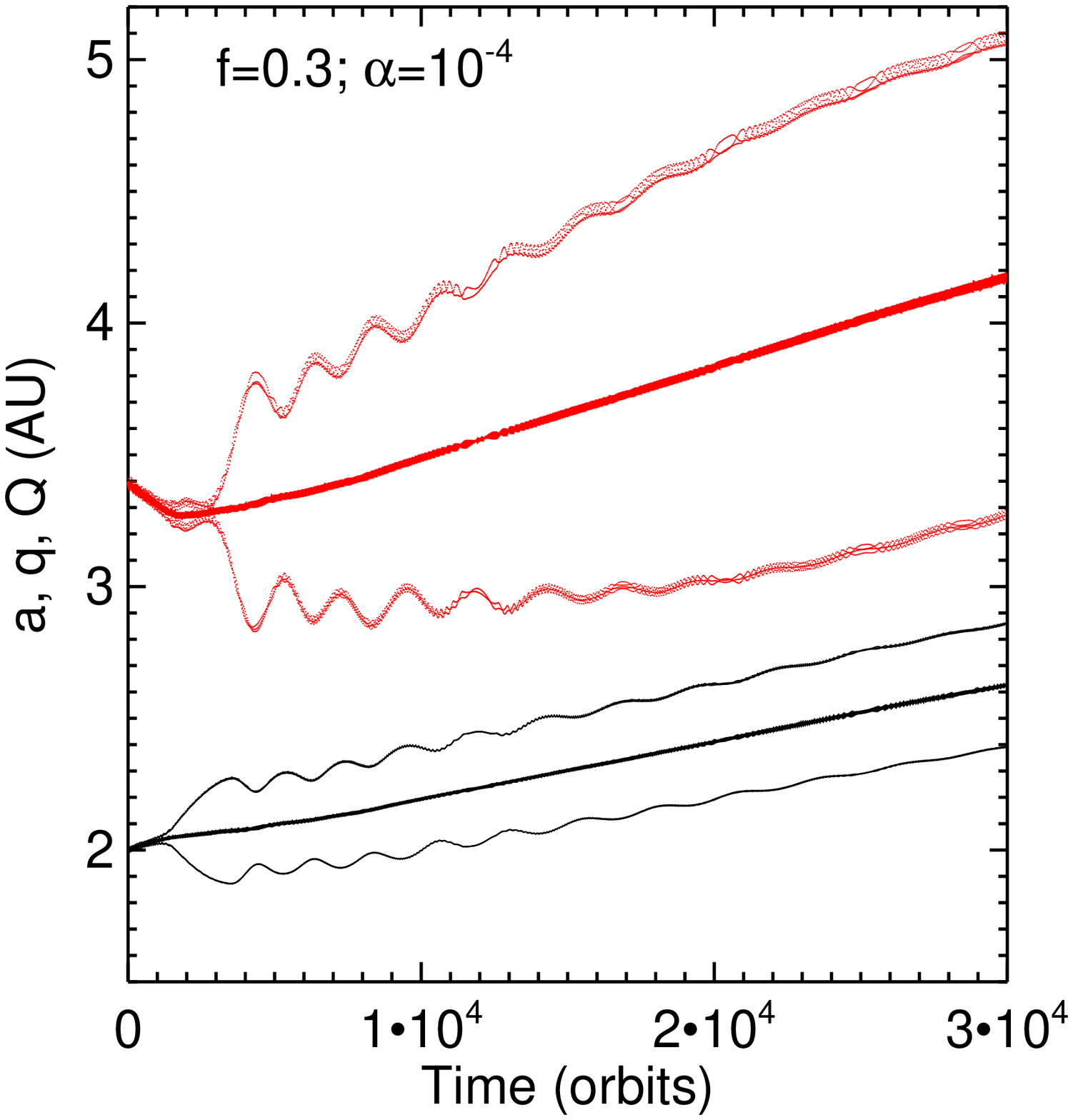}
\includegraphics[width=0.33\textwidth]{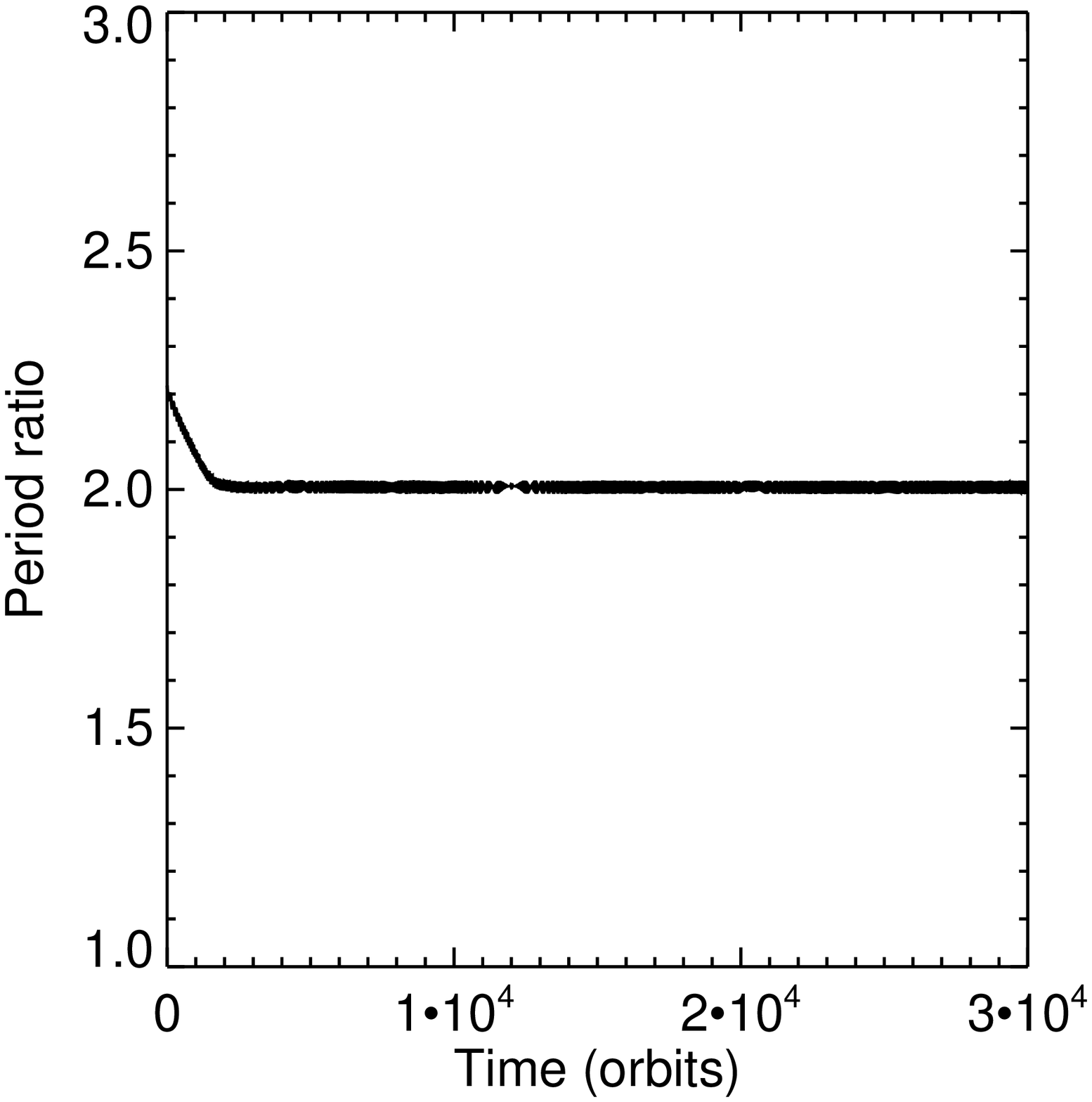}
\includegraphics[width=0.33\textwidth]{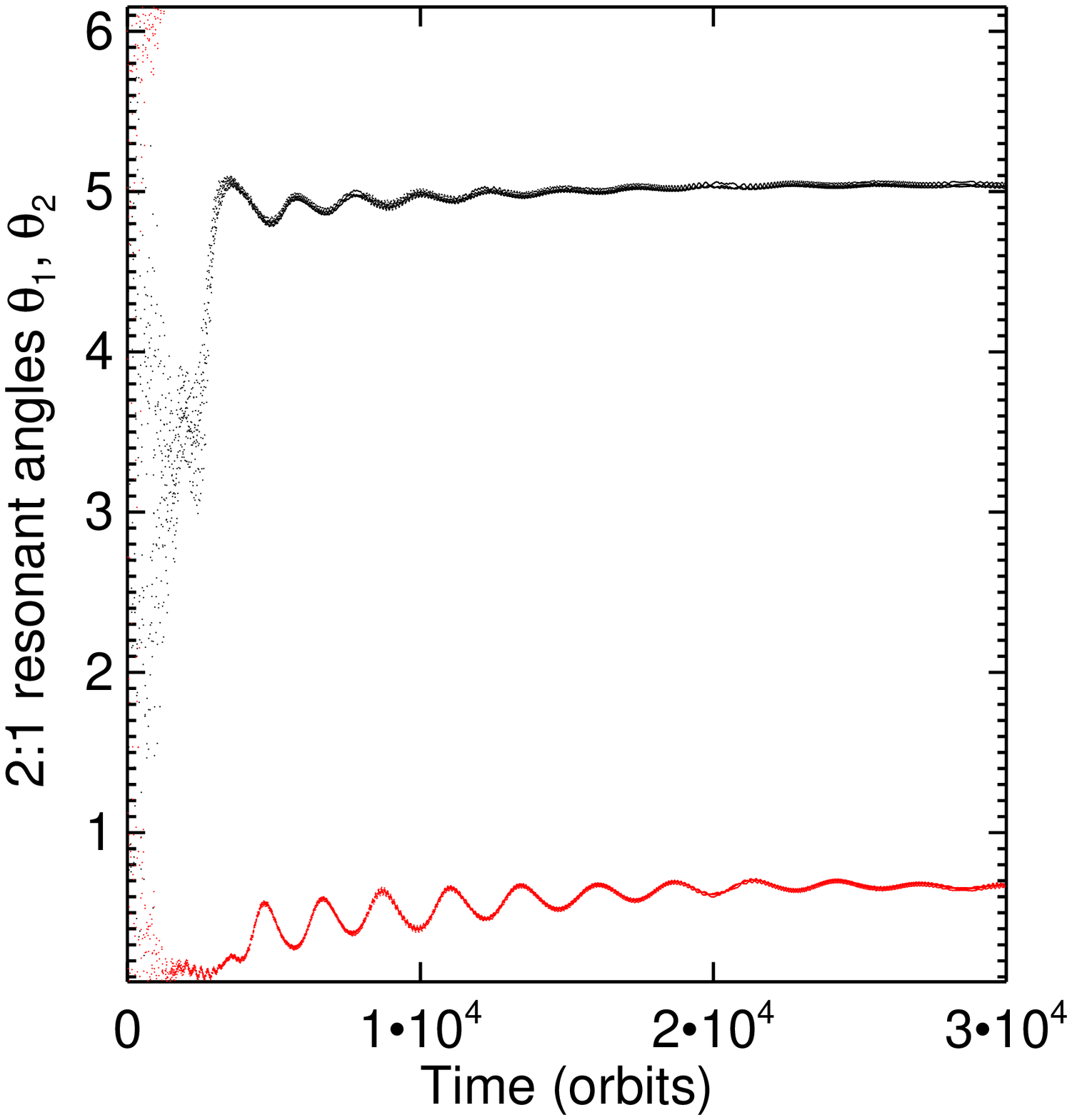}
\includegraphics[width=0.33\textwidth]{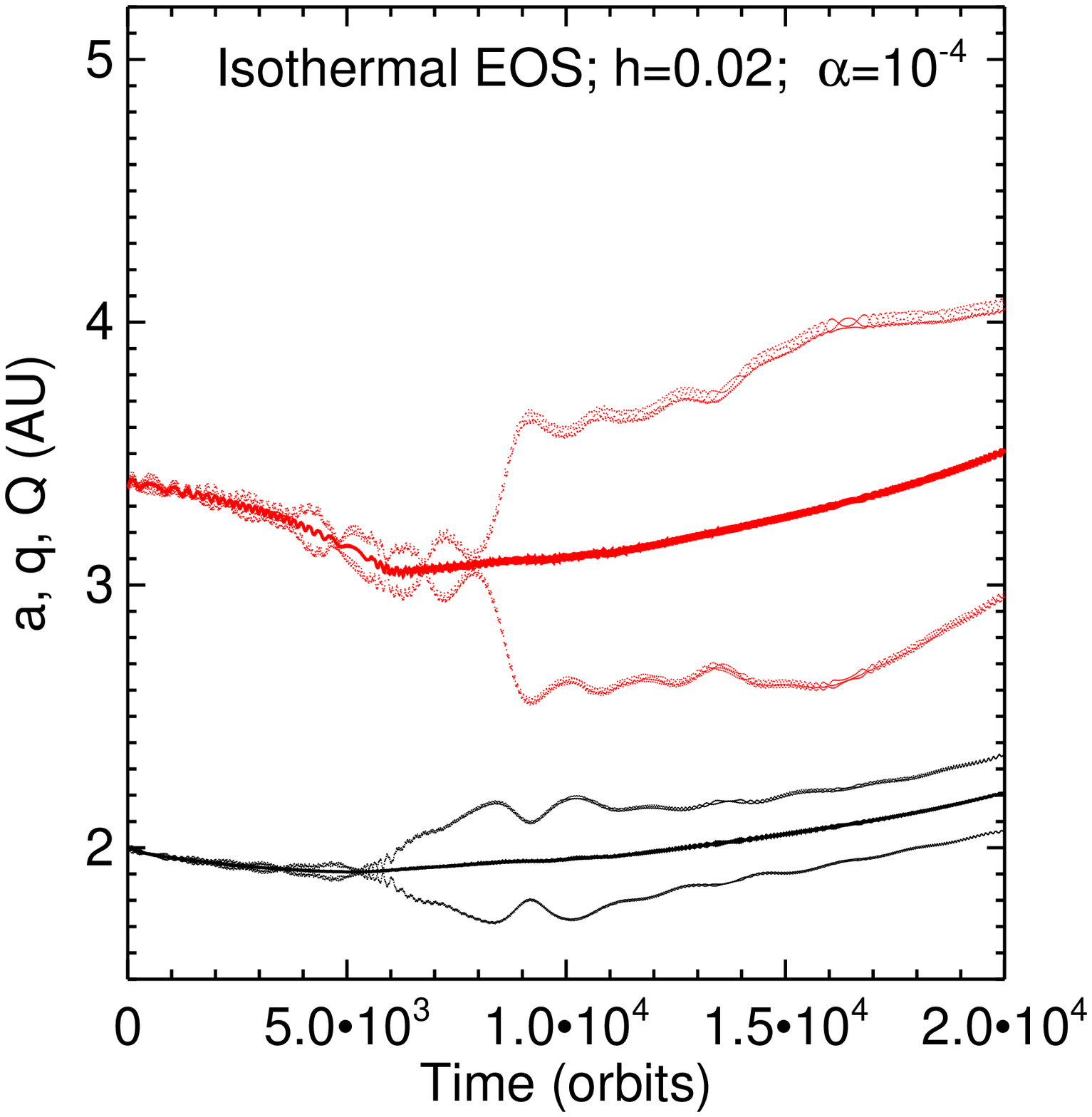}
\includegraphics[width=0.33\textwidth]{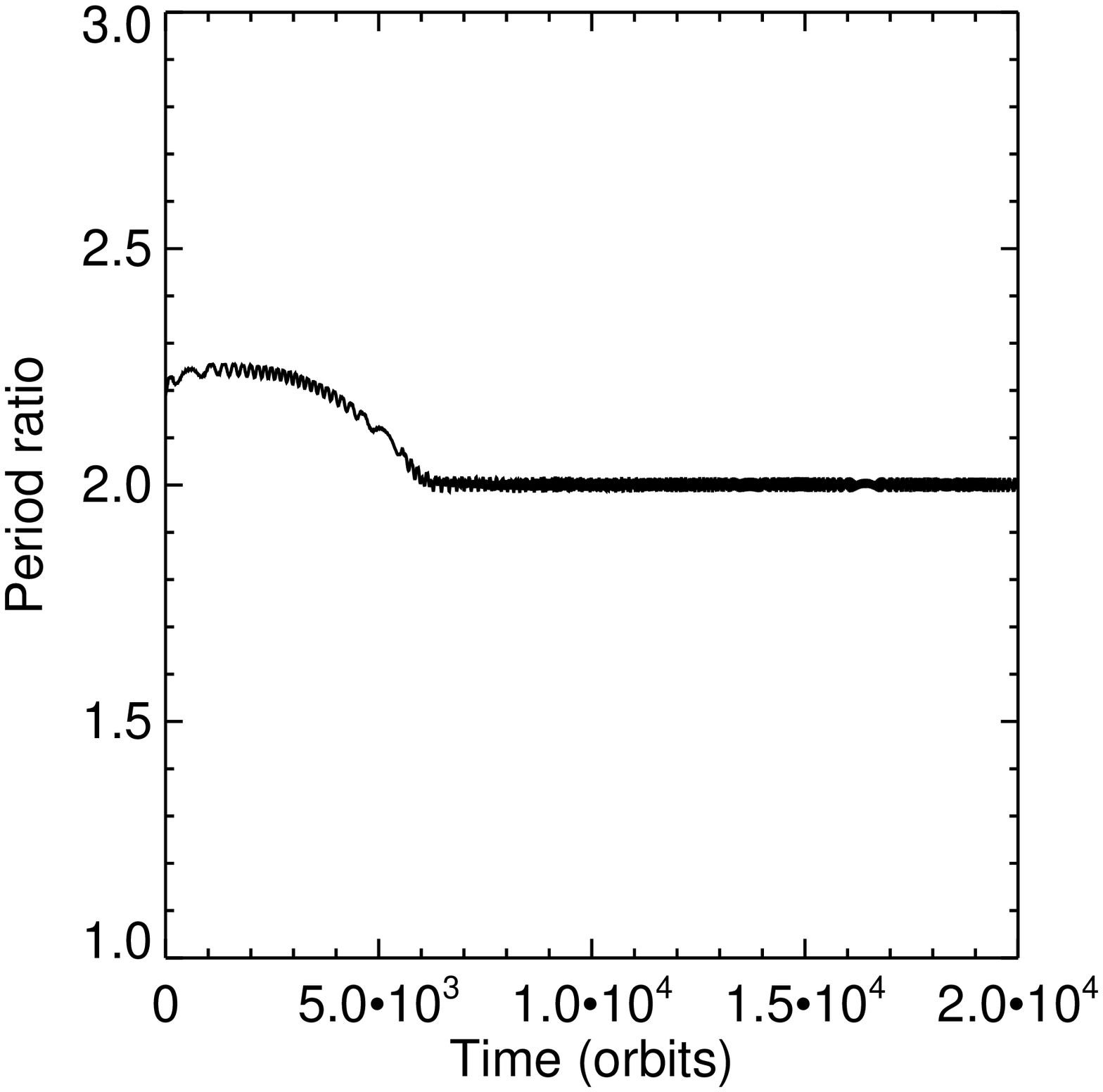}
\includegraphics[width=0.33\textwidth]{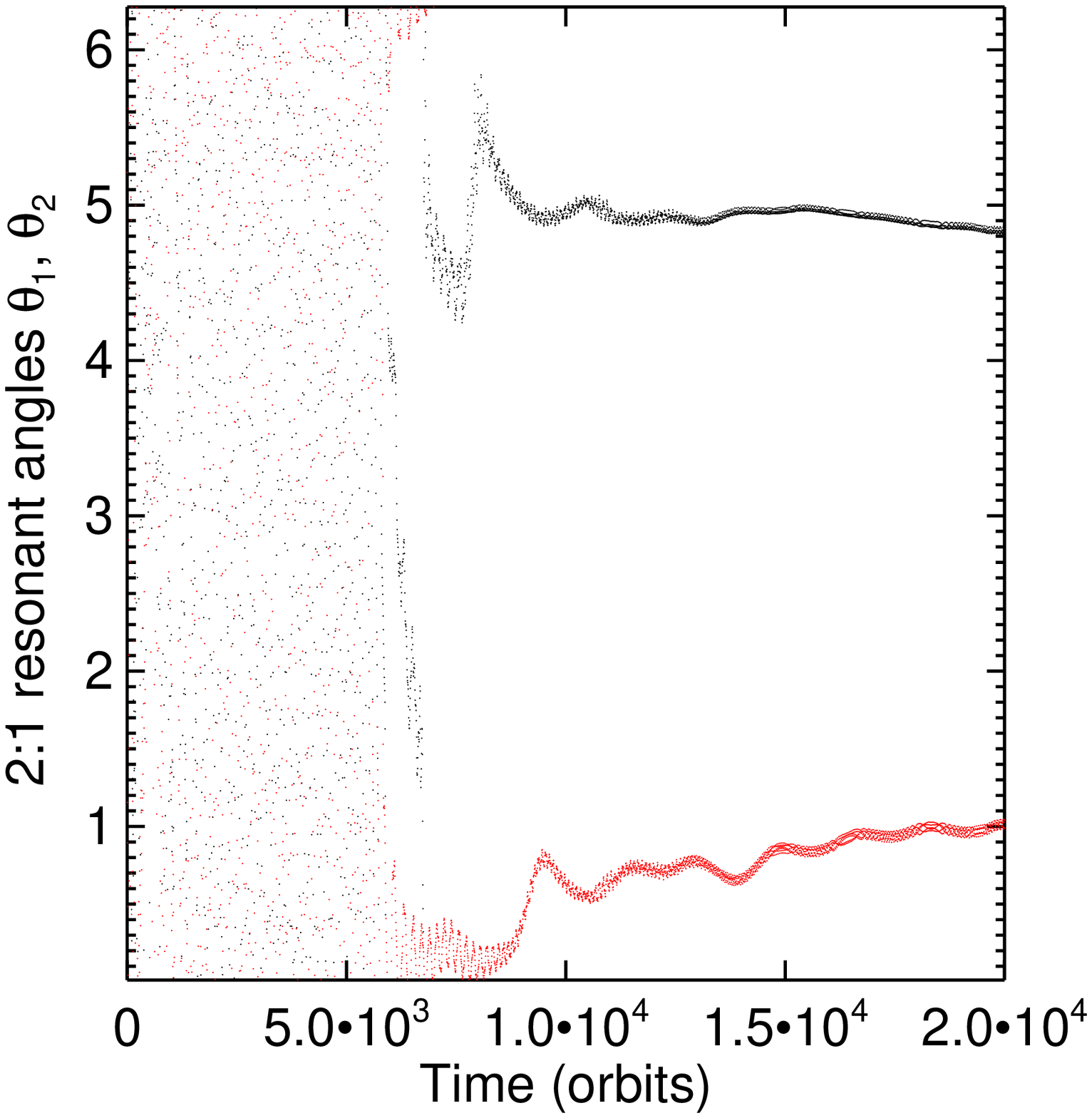}
\caption{{\it Upper panel:}  time evolution of the planets' semi-major axes $a$, perihelia $q$ and aphelia $Q$ 
(left) for a disk model with $f=3$ and $\alpha=10^{-4}$. The middle panel 
shows the evolution 
of the period ratio and the right panel the evolution of the resonant angles 
$\theta_1=3\lambda_S-2\lambda_J-\varpi_S$ (black) and $\theta_2=3\lambda_S-2\lambda_J-\varpi_J$ (red)
associated with the 3:2 resonance, where $\lambda_J$ ($\lambda_S$) and $\varpi_J$ ($\varpi_S$) are the 
longitude and pericenter of Jupiter (Saturn).
{\it Middle panel:} Same but for a disk model with $f=0.3$ and $\alpha=10^{-4}$. The middle right panel shows the evolution of the resonant angles 
$\theta_1=2\lambda_S-\lambda_J-\varpi_S$ (black) and $\theta_2=2\lambda_S-\lambda_J-\varpi_J$ (red)
associated with the 2:1 resonance. 
{\it Lower panel:} Same but for an isothermal run with $\alpha=10^{-4}$ and $h=0.02$.}
\label{fig:at}
\end{figure*}

\begin{figure}
\includegraphics[width=\columnwidth]{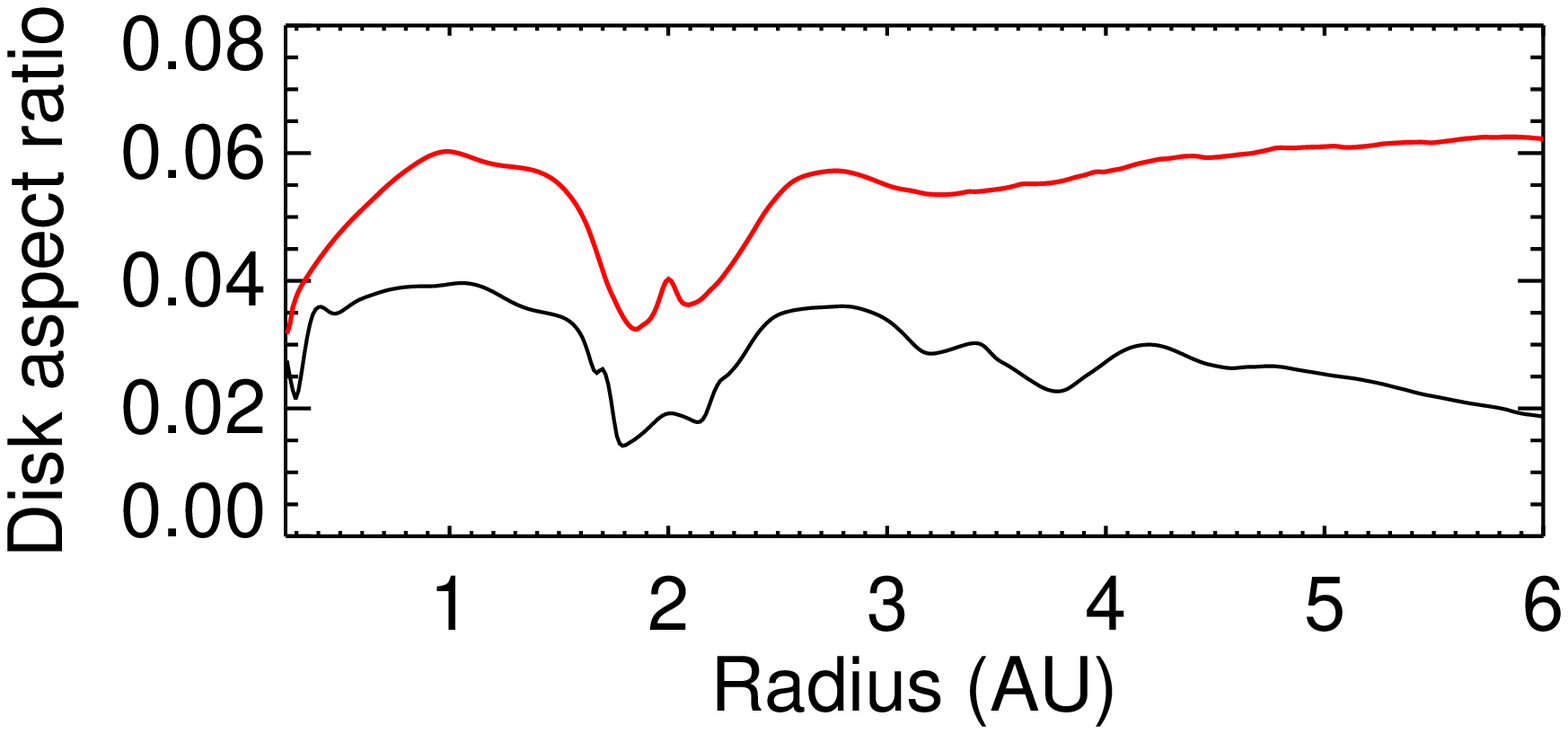}
\includegraphics[width=\columnwidth]{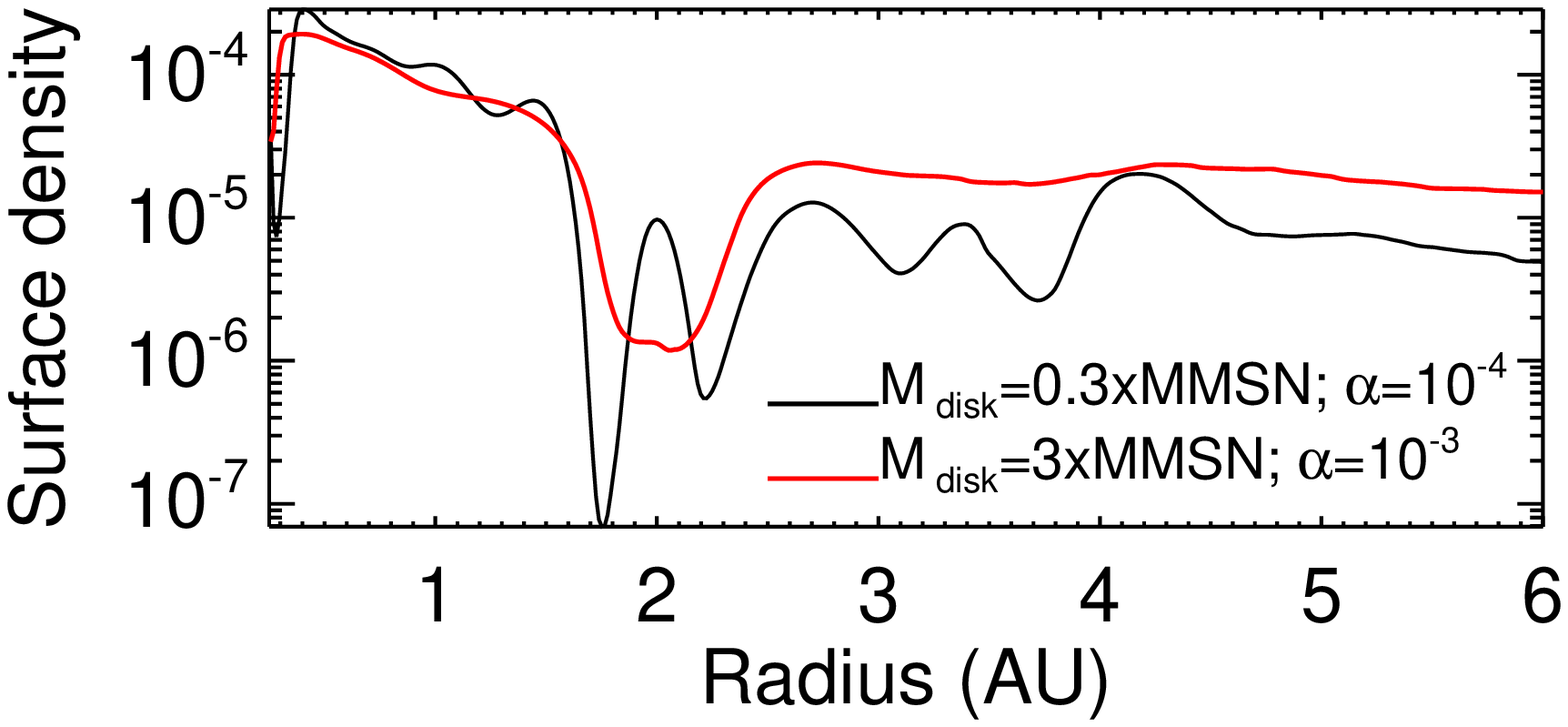}
\caption{{\it Upper panel:} Disk aspect ratio as a function of radius for the models with $f=0.3$, $\alpha=10^{-4}$ (black) and $f=3$, 
$\alpha=10^{-3}$ (red). Here Jupiter and Saturn are held on circular orbits with $a_J=2$ and $a_S=3.4$ AU. 
{\it Lower panel:} Surface density profile  for the same models.}
\label{fig:dh}
\end{figure}

\begin{figure*} 
\centering
\includegraphics[width=0.33\textwidth]{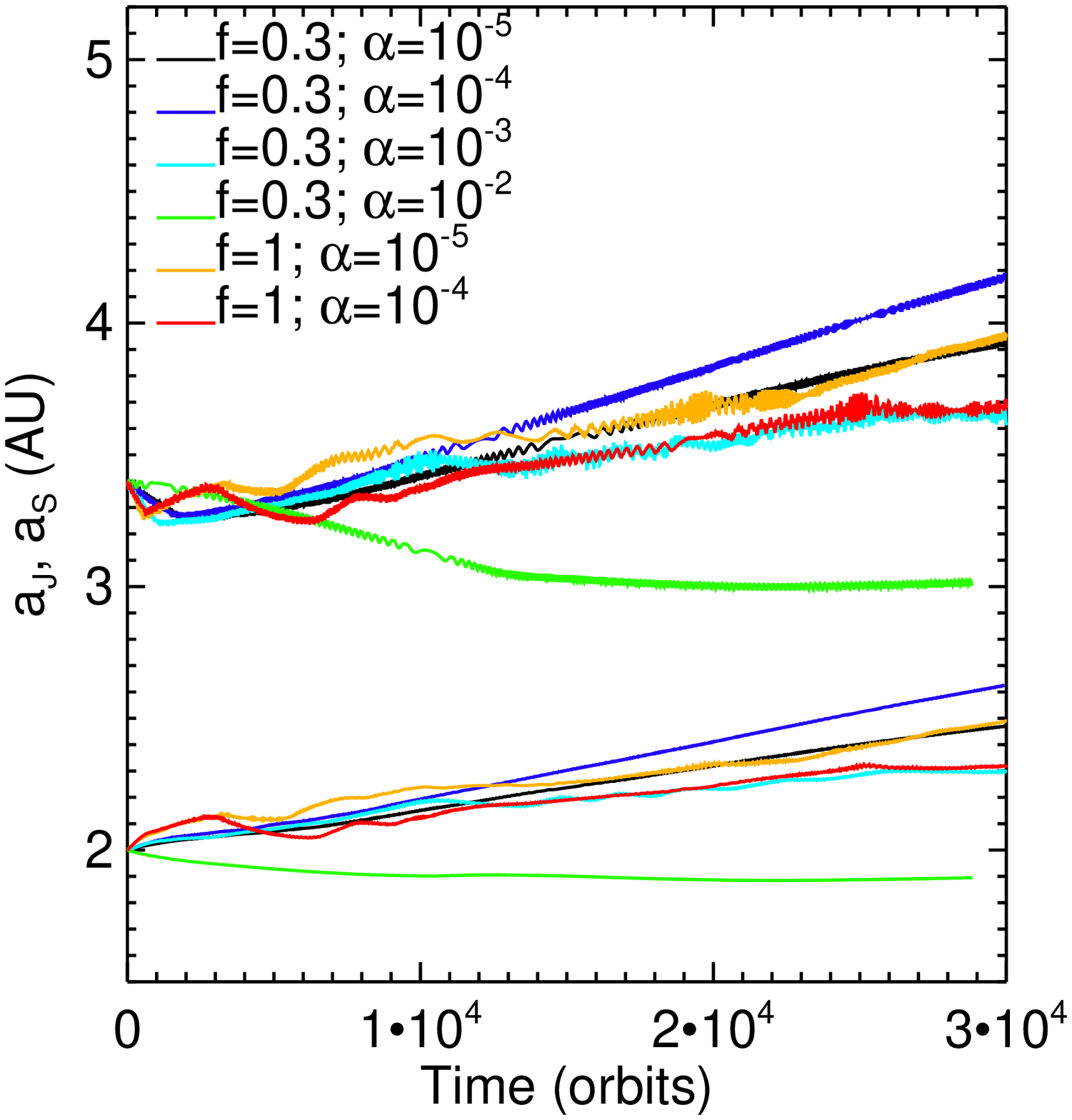}
\includegraphics[width=0.33\textwidth]{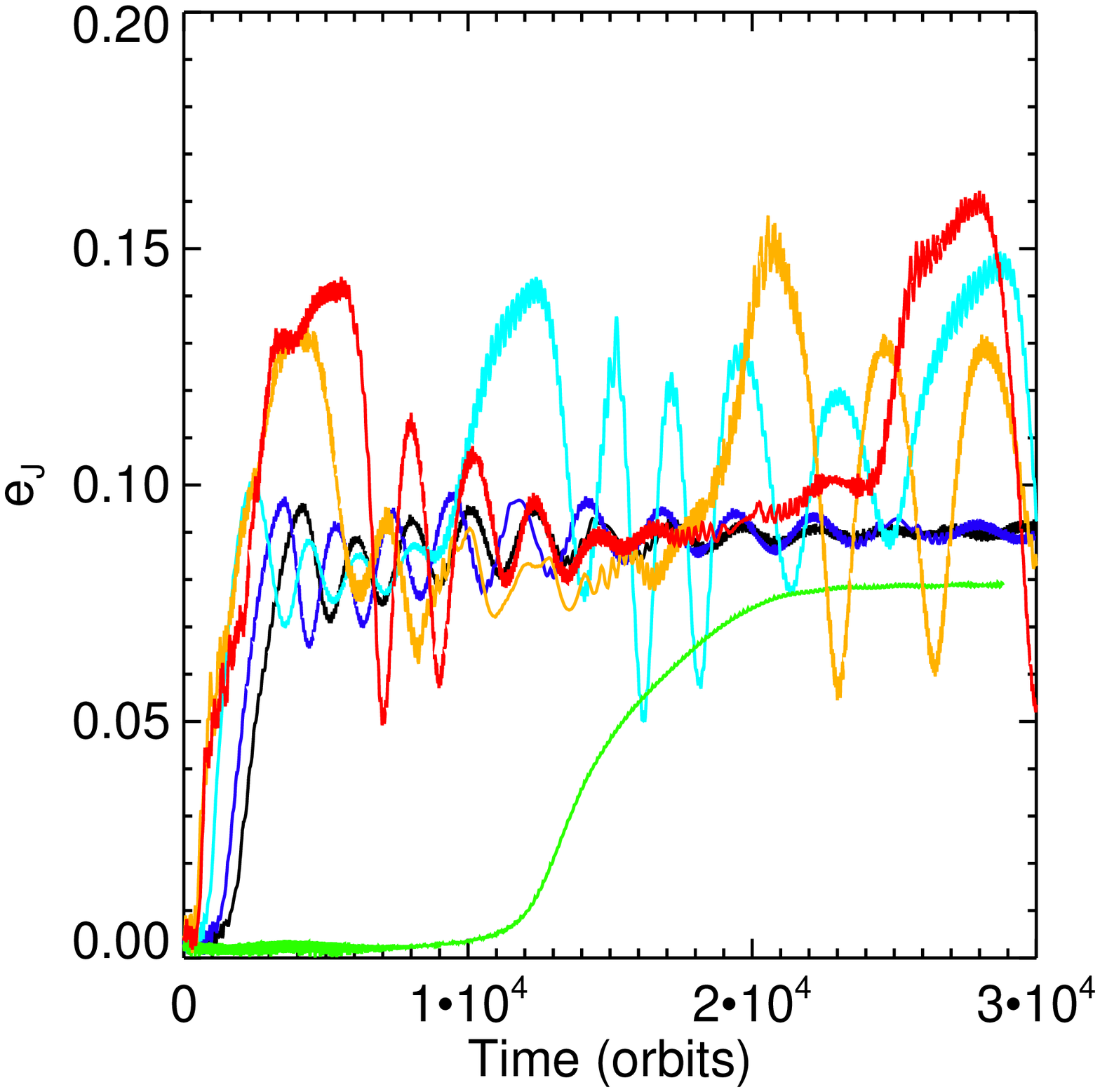}
\includegraphics[width=0.33\textwidth]{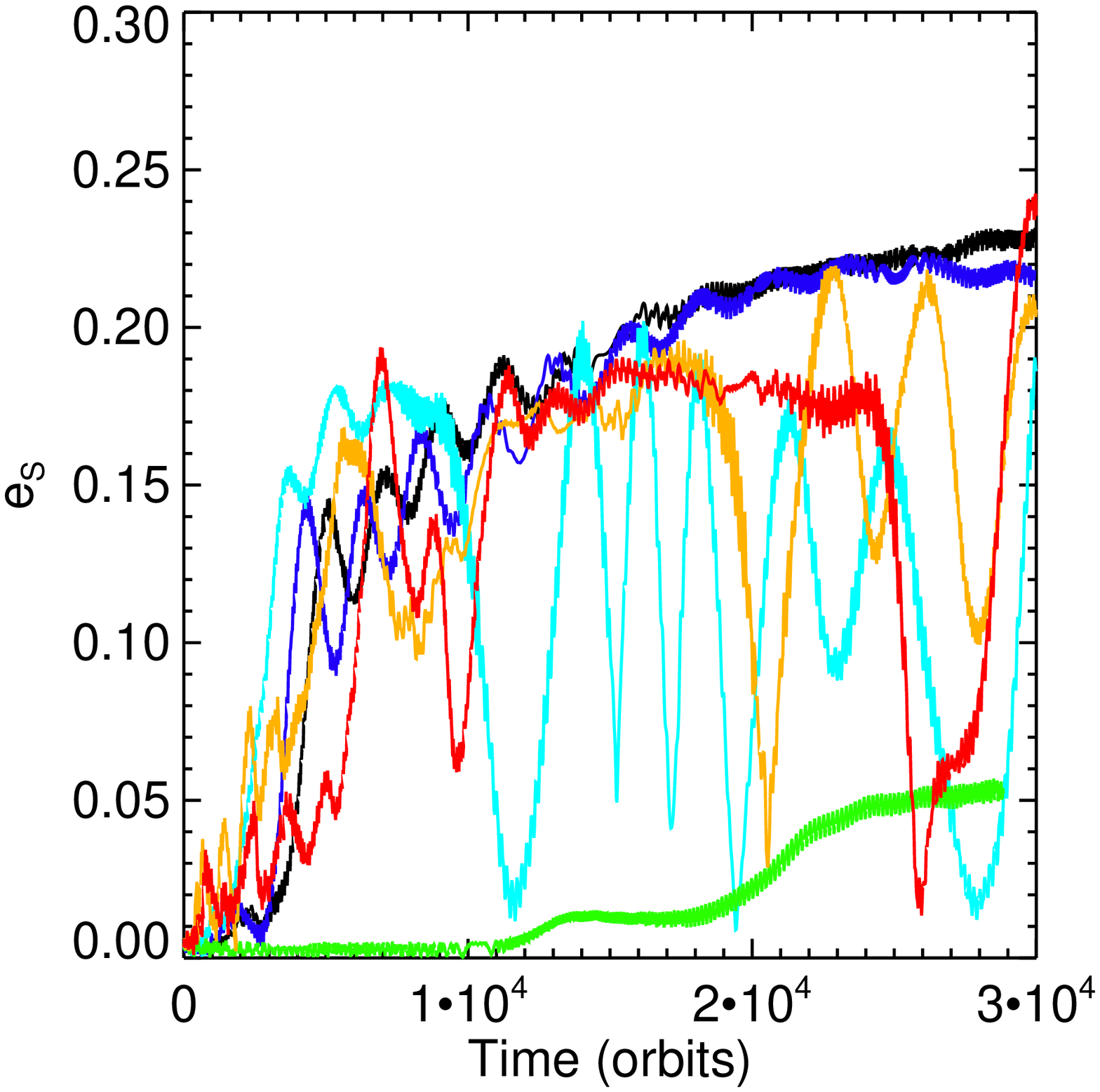}
\caption{Time evolution of Jupiter and Saturn semi-major axes (left panel), Jupiter's eccentricity (middle panel) and 
Saturn's eccentricity (right) for simulations that led to the formation of a stable 2:1 resonance.}
\label{fig:models21}
\end{figure*}

\begin{figure}
\centering
\includegraphics[width=0.8\columnwidth]{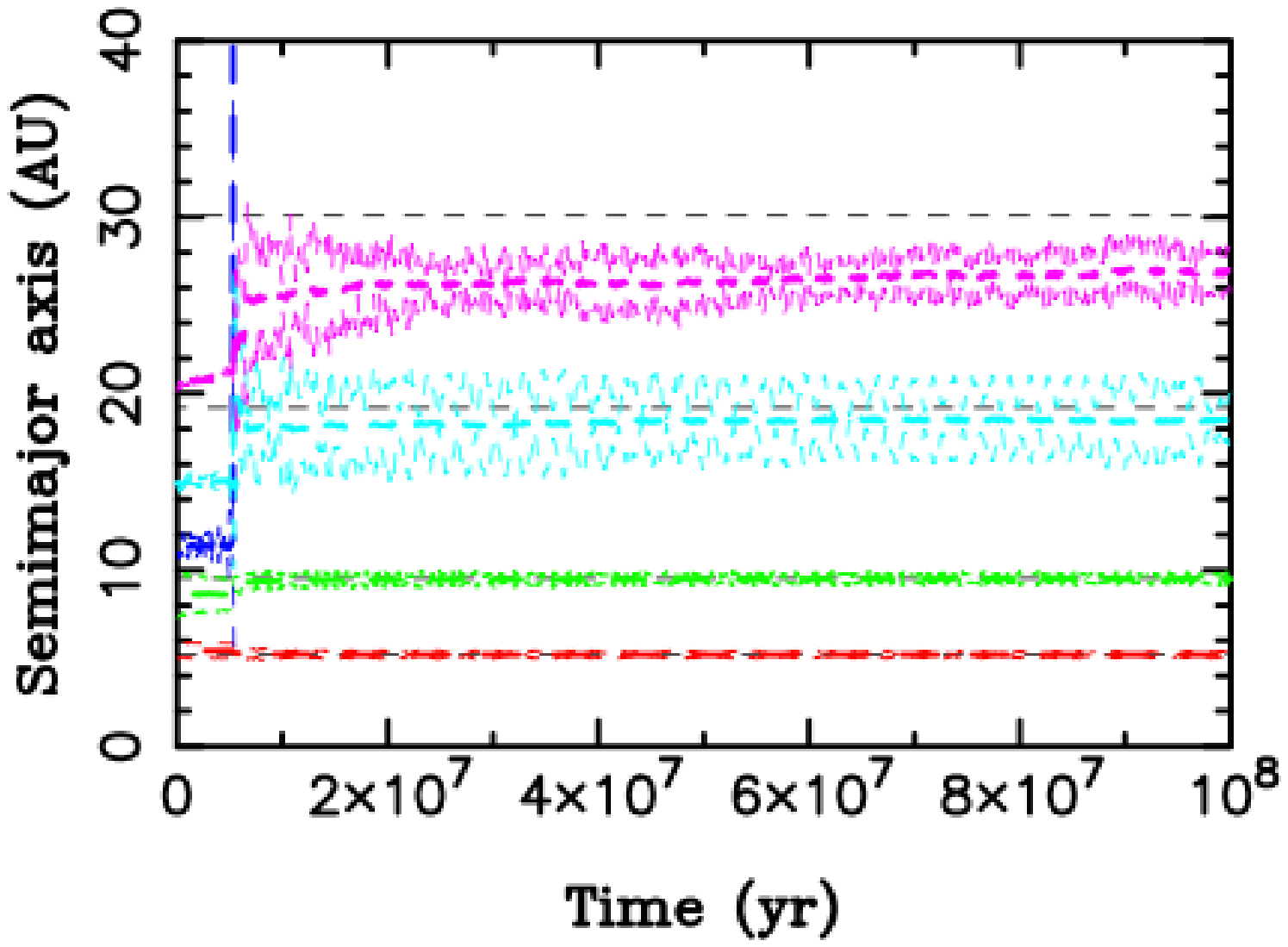}
\includegraphics[width=0.8\columnwidth]{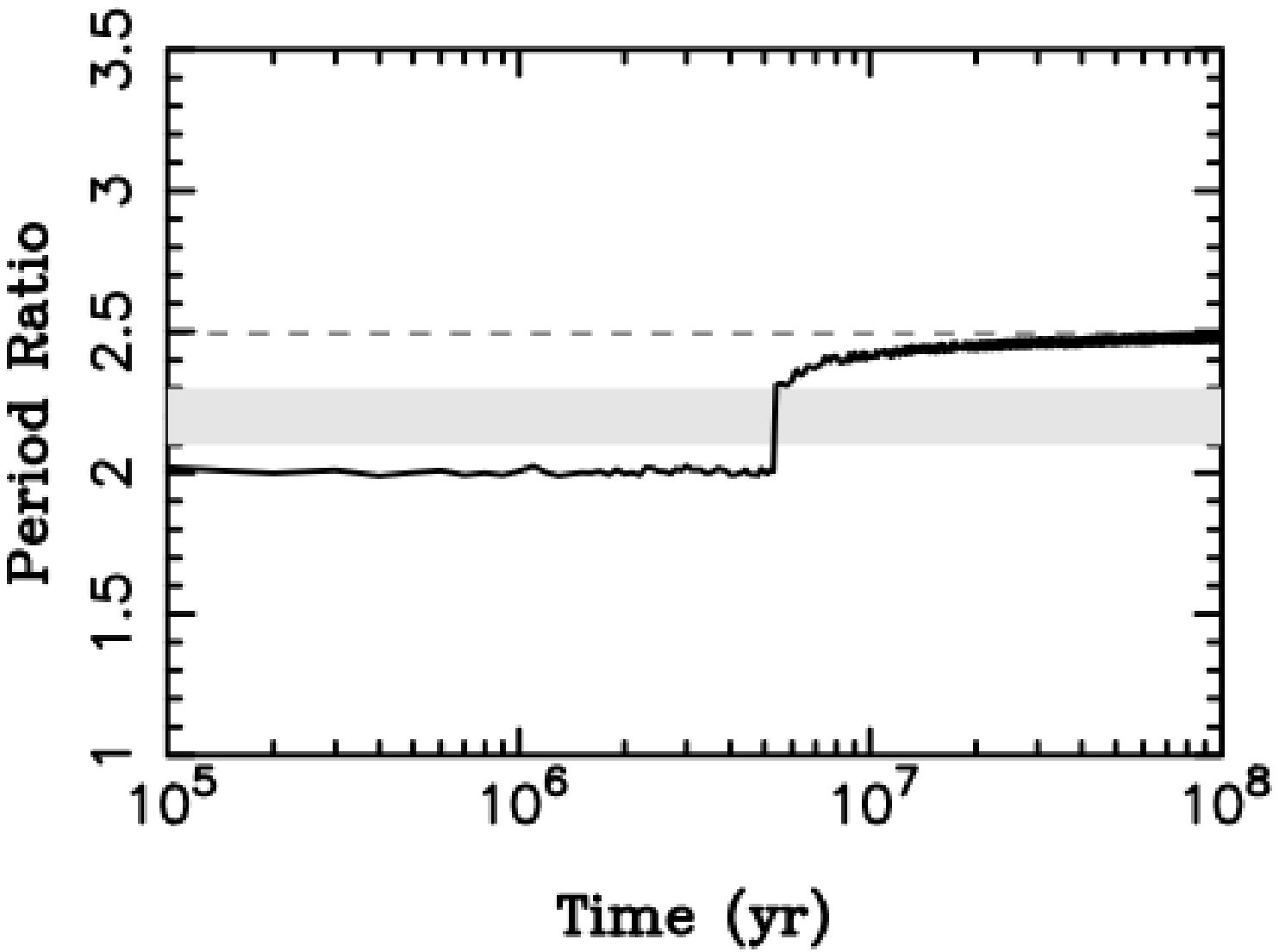}
\caption{{\it Upper panel:} Time evolution of planets' semi major axes for a five-planet Nice model run 
with initial configurations resulting from the disk model with $f=0.3$, $\alpha=10^{-4}$. The five planets 
were started in a (2:1, 3:2, 3:2, 3:2) resonant chain and the mass of the planetesimal disk is $m_{disk}=20$ $M_\oplus$. 
{\it Lower panel:} Period ratio between between Jupiter and Saturn as a function of time.}
\label{fig:nice}
\end{figure}

\section{Hydrodynamical model}

Simulations were performed using the GENESIS (De Val-Borro et al. 2006) numerical code that solves the equations governing the disk evolution on a polar grid $(R,\phi)$. The code's energy equation reads:
\begin{equation}
\frac{\partial e}{\partial t}+\nabla \cdot (e{\bf v})=-p(\nabla\cdot{\bf v})+Q^+_{visc}-Q^-_{rad}
\label{eq:energy}
\end{equation}
 where $\bf v$ is the gas velocity, $e$ the  thermal energy density, $\gamma$ the adiabatic index (set to $\gamma=1.4$). 
$p=(\gamma -1) e$ is the pressure which is related to the disk temperature $T$ and surface density $\Sigma$ as 
 $p={\cal R} \Sigma T/\mu$, where ${\cal R}$ is the ideal gas constant and $\mu=2.35$ the mean molecular weight. $Q^+_{visc}$ is  the viscous heating term where viscous stresses are modeled using the standard 'alpha' prescription for the disk viscosity 
$\nu=\alpha c_s H$, where $c_s=(\gamma  {\cal R} T/\mu)^{1/2}$ is the sound speed and $H$ the disk scale height which is 
related to the angular velocity $\Omega$ and the isothermal sound speed $c_{s,iso}=c_s/\sqrt{\gamma}$ as 
$H=c_{s,iso}/\Omega$. 
$Q^-_{rad}=2\sigma_B T_{eff}^4$ is the local radiative cooling term, 
where $\sigma_B$ is the Stephan-Boltzmann constant and $T_{eff}$ is the effective temperature which 
is computed using the opacity law of Bell \& Lin (1994).  In this work, effects resulting from stellar heating 
are not considered in the energy budget.


We employ $N_R=896$ radial grid cells uniformly distributed between $R_{in}=0.25$ and $R_{out}=12$ AU, and $N_\phi=700$ azimuthal grid cells. At the inner edge, we use a viscous outflow boundary condition (see Pierens \& Nelson 2008). At the outer edge, we employ a wave-killing zone for $R>11.4$ to avoid wave reflections.

 The initial surface density profile is 
$\Sigma=f\;\Sigma_{MMSN} ( R/1 AU)^{-3/2}$,
 where 
$\Sigma_{MMSN}=2\times 10^{-4}$ in dimensionless units is the surface density at $1$ AU of the MMSN, and  $f$ is 
an enhancement factor.  The initial temperature profile is such that $T\propto R^{-1}$. Due to the action of source terms in Eq. $1$, however, this initial temperature quickly evolves until an equilibrium state is reached. 
 
Jupiter and Saturn initially evolve on circular orbits with semi-major axes $a_J=2$ and $a_S=3.4$ AU respectively, just exterior to their mutual $2:1$ resonance. To give them sufficient time to open a gap, the planets are held on fixed circular orbits for $\sim 500$  orbits, and then are released   and evolve under the action of disk torques.  When calculating the disk torques, we exclude the material contained within a distance $0.6R_H$ from the planets
  using a Heaviside filter (Crida et al. 2009) , where $R_H$ is the Hill radius.  

In our simulations we tested the effect of two parameters: the disk's  surface density  and viscosity.  We tested $\alpha=10^{-5}, 10^{-4},10^{-3}, 10^{-2}$, and for each value of $\alpha$ we performed four different simulations with $f=0.3, 1, 3, 10$, for a total of 16 simulations.  

\section{Results}

The results of our simulations are illustrated in Fig. \ref{fig:bilan}. There are four qualitatively different outcomes 
 shown with different symbols. Runs that are labelled "3:2 resonance"  (downwards triangles) produced  outward migration of Jupiter and Saturn in a 3:2 resonance (Masset \& Snellgrove 2001; Morbidelli \& Crida 2007). The top panels of Fig. \ref{fig:at} show one such outcome, for the simulation with $\alpha=10^{-4}$ and $f=3$.  The planets migrate convergently and become trapped in a 2:1 resonance at $t\sim 100$ orbits (see top middle panel of Fig. \ref{fig:at}). Due to Saturn's fast migration, the planets break free from the 2:1 resonance and become trapped in a $3:2$ resonance at $t\sim 300 $ orbits. From that time on, Jupiter and Saturn evolve in a common gap and the 
resulting change in the torque balance, together with the flux of gas from the outer disk across the gap, makes them migrate outward.  The 3:2 resonance is maintained throughout (resonant angles shown in the top right panel of Fig. \ref{fig:at}).

The  middle panels of Fig. \ref{fig:at} show the evolution of a simulation with $f=0.3$ and $\alpha = 10^{-4}$.  Jupiter and Saturn were captured in 2:1 resonance.  This triggered outward migration that was maintained to the end of the simulation.  As for the 3:2 resonance, outward migration was maintained by a continuous flux of gas across the planets' common gap. This flux acts to replenish the inner disk and also to exert a positive corotation torque on the planets.  The main difference between the two simulations illustrated in Fig.~\ref{fig:at} is that the planets' eccentricities are much higher migrating outward in 2:1 resonance.  This occurs because the gap is wider and deeper in the simulation with the gas giants in 2:1 resonance, resulting in a weaker damping of the planets' eccentricities.  

Jupiter and Saturn should be caught in stable 2:1 resonance when Saturn's migration is slower than a critical rate.  The critical velocity for capture in the 2:1 resonance $\sim (M_J/M_*)^{4/3} a_S \Omega_S$ (D'Angelo \& Marzari 2012), where $\Omega_S$ is the Saturn's angular velocity. Assuming an  isothermal type I migration rate for Saturn (Tanaka et al. 2002) and that Jupiter does not migrate, D'Angelo \& Marzari (2012) found that Jupiter and Saturn would become trapped in 2:1 resonance if the surface density at $2$ AU is $\lesssim 650$ $g.cm^{-2}$.  Although we use a non-isothermal 
disk model here, Fig. \ref{fig:bilan} shows that our simulations are consistent with this estimate, since all runs with $f=0.3$ formed a stable 2:1 resonance. An alternative possibility,  which is the one that actually happens in the 
simulations, is that both the disk aspect ratio and viscosity are small enough for Saturn undergo a slow Type II migration (Lin \& Papaloizou 1993). In that case, capture in 2:1 resonance is expected provided that the condition for gap opening is satisfied (Crida et al. 2006):
\begin{equation}
1.1\left(\frac{q_S}{h^3}\right)^{-1/3}+\frac{50\nu}{q_Sa_S^2\Omega_S}<1
\end{equation}
where $q_S=M_S/M_*$ is the Saturn mass ratio. For a thin disk with $h=0.03$, $\alpha \lesssim 3\times 10^{-3}$ whereas 
for $h=0.05$, capture in the 2:1 resonance should arise for $\alpha \lesssim 4\times 10^{-4}$. This estimate also agrees with our simulations since runs with disc masses corresponding to the MMSN and $\alpha \leq 10^{-4}$ produced a stable 2:1 resonance.  

When viscous stresses are the only heating process, a low disk mass (small value of $f$) and a modest viscosity (small $\alpha$) produces a thin disk. Figure \ref{fig:dh} (top panel) shows the aspect ratio as a function of radius for two disks; a low-mass, low-viscosity disk with $f=0.3$, $\alpha=10^{-4}$ and a high-mass, modest-viscosity disk with $f=3$, $\alpha=10^{-3}$. Here, Jupiter and Saturn are held on circular orbits at $2$ and $3.4$ AU, respectively (exterior to 2:1 resonance). In the high-mass disk the aspect ratio is $h\sim 0.04$ ($h\sim 0.05$) at the location of Jupiter (Saturn). In contrast, in the low-mass disk $h\sim 0.02$ ($h\sim 0.03$) at Jupiter's (Saturn's) orbit. For the low-mass, low-viscosity disk, the gaps opened by the planets are wider and deeper than in the high-mass disk (see bottom panel of Fig. \ref{fig:dh}).  In the low-mass, low-viscosity disk the gas near Saturn's orbit is significantly depleted, imparting a strong positive torque on Jupiter which may favor outward migration (Morbidelli \& Crida 2007).

 A small aspect ratio thus appears to be the key factor in causing outward migration with Jupiter and Saturn in 2:1 resonance. In principle, outward migration should also occur in isothermal disk models with small aspect ratios.  To test 
this hypothesis,  we performed a simulation using an isothermal equation of state with $h=0.02$ and  $\alpha=10^{-4}$ 
and which
indeed resulted in outward migration with Jupiter and Saturn in a 2:1 resonance, as illustrated in the lower panel of 
Fig. \ref{fig:at}.

 Fig. \ref{fig:models21} shows the results of radiative simulations that produced stable 2:1 resonances.  For a MMSN model, the condition for outward migration is $\alpha\lesssim 10^{-4}$ whereas for $f=0.3$, outward migration occurs for $\alpha \lesssim 10^{-3}$. This is consistent with the expectation that the aspect ratio in radiative disks should be a function of  the 
product $ \nu \Sigma$  (Bitsch et al. 2014).  Outward migration tends to be faster for smaller $f$ and smaller $\alpha$. The run with $f=0.3$ and $\alpha=10^{-5}$, which exhibits a smaller outward migration rate than in the case with $\alpha=10^{-4}$, is an exception; this probably occurs because  the inner edge of Jupiter's gap is located further away from the planet in that case.

\section{Effect of stellar irradiation}
  Results from the provious section suggest that the 2:1 resonance can lead to outward migration in disks with very small aspect ratios $h\sim 0.02$.  For a radiative disk subject to viscous heating only, low values of the aspect ratio correspond to small disk masses and/or viscosities and therefore to low accretion rates.  In that case, stellar irradiation may contribute significantly to the disk temperature structure and possibly dominate over viscous heating, but this depends strongly on the disk 
metallicity (Bitsch et al. 2014). Recent hydrodynamical simulations of 
 the structure of disks with constant accretion rate (Bitsch et al. 2014) show that  in the late stages of evolution, a metallicity of $0.05$ can lead to an aspect ratio $h\sim 0.03$ in the inner parts,  and it is expected that even smaller values for $h$ to 
 be reached for lower metallicities. Moreover, 1D modeling  of viscous protoplanetary disks subject to 
 stellar irradiation (Bailli\'e \& Charnoz, in prep.) also  shows  that at very late times, the aspect ratio can be typically $h\sim 0.02$ at $5$ AU. Both of these studies therefore suggest that the conditions for Jupiter and Saturn to migrate outward 
 in 2:1 resonance may be  fulfilled under certain conditions. It remains to be seen, however,  whether outward migration of Jupiter and Saturn in 2:1 resonance triggered in a part of the disk with a low aspect ratio could be maintained if the planets migrated into an outer, thicker part of the disk. Clearly, more sophisticated hydrodynamical simulations are needed to definitely assess the effect of stellar heating on these results.

\section{Evolution after gas disk dispersal}

We now turn our attention to the later evolution of the outer Solar System.  Our goal is to test whether simulations showing outward migration of Jupiter and Saturn in 2:1 resonance are compatible with the current architecture of the Solar System. 

We selected three disk models that produced outward migration in 2:1 resonance at different rates (with $f=0.3$, $\alpha=10^{-4}$; $f=0.3$, $\alpha=10^{-3}$ and $f=1$, $\alpha=10^{-5}$).  We restarted each simulation and artificially dissipated the disk by forcing the gas surface density to decay exponentially with an e-folding time $t_{dis}=10^4$ yr. We then used the outputs of these runs, rescaled such that the initial semi-major 
axis of Jupiter was $\sim  5$ AU, as initial conditions for N-body simulations of the evolution of the outer Solar System after dispersion of the gas disk. 

Following Nesvorny (2011) and Nesvorny \& Morbidelli (2012), we performed simulations with Uranus and Neptune initially located in a resonant chain with the giant planets.  The resonant chain is generated using both hydrodynamic 
and N-body simulations to identify the resonant configurations that are compatible with Jupiter 
and Saturn in 2:1 resonance.  We also considered a five-planet case where an additional planet with mass comparable to that of Uranus/ Neptune was placed between the orbits of Saturn and Uranus.  In each case, the orbit of one planet was purposely shifted by $180^\circ$ in mean anomaly to artificially disrupt the resonant chain and trigger an instability (see Levison et al 2011). 
An outer planetesimal disk was included with masses of $m_{disk}=20, 35, 50, 100 \;M_\oplus$. For each case, we performed $30$ simulations with different, randomly-generated,  radial ranges and surface density profiles of the  planetesimal disk. 

In the four-planet case,  the initial states deduced from our three disk models did not produce good Solar System analogs. Low disk masses typically led to final systems with fewer than four planets. For high disk masses the ice giants migrated too far and the giant planets end up on too-circular orbits due to too strong disk-induced eccentricity damping (Nesvorny 2011). We obtained better results in simulations with an extra ice giant.  This is broadly consistent with Nesvorny \& Morbidelli (2012), who found a much higher success rate in reproducing the current architecture of the Solar System in the five-planet case.  This was true for Jupiter and Saturn in both 3:2 and 2:1 resonance.  
 
Figure \ref{fig:nice} shows the results of a successful five-planet run with initial configuration from the model with $f=0.3$, $\alpha=10^{-4}$. The planets were started in a (2:1, 3:2, 3:2, 3:2) resonant chain and the planetesimal disk mass was $m_{disk}=20$ $M_\oplus$. The extra ice giant (in blue) was ejected from the system after $\sim 6$ Myr, and the subsequent migration of the  giant planets was marginal.  This run satisfies most of the constraints from Nesvorny \& Morbidelli (2012).  The final semi major axes of the planets are $5.20$, $9.33$, $18.01$ and $28.40$ AU and their final eccentricities are $0.029$, $0.045$, $0.058$ and $0.03$.  Neptune is a little too close to the Sun and Jupiter's eccentricity is slightly smaller than its current value, but  the lower panel of Fig. \ref{fig:nice} shows that the final Jupiter-Saturn period ratio is almost perfect.  When the ice giant is ejected at $\sim 6$ Myr, the Jupiter-Saturn period ratio jumps from $(a_S/a_J)^{1.5}/<2.1$ to  $(a_S/a_J)^{1.5}/> 2.3$ in $< 1$ Myr such that the $\nu_5$ secular resonance jumps, rather than sweeps, across the inner Solar System (Brasser et al 2009; Morbidelli et al 2010; called constraint D by Nesvorny 2011 and Nesvorny \& Morbidelli 2012).  In this example the secular amplitude ${\it e_{55}}$ is also close to the current value.

Although we did not perform a broad statistical study as in Nesvorny \& Morbidelli (2012), the existence of successful cases such as the one from Fig.~\ref{fig:nice} shows that the outward migration of Jupiter and Saturn in 2:1 resonance is consistent with the late evolution of the outer Solar System.

\section{Discussion and Conclusion}



We have shown that a Grand Tack can occur with Jupiter and Saturn in 2:1 resonance for a limited range of the parameter space defined by the disk's mass and viscosity.  Outward migration in 2:1 takes place in relatively low-mass ($M_{disk} \lesssim M_{MMSN}$), low-viscosity ($\alpha \lesssim 10^{-3}$) disks  that tend to have very small aspect ratios ($h \sim 0.02$). Compared with outward migration in 3:2 resonance, the biggest difference in the evolution when Jupiter and Saturn are in 2:1 resonance is that their eccentricities are higher, with $e_{J,S} \approx$0.05-0.2.  

Outward resonance of Jupiter and Saturn in 2:1 resonance is consistent with a later, Nice model instability.  We used the simulations in which the gas giants migrated outward in 2:1 resonance as inputs to N-body simulations of the evolution of the outer Solar System after dispersal of the gas disk. Instabilities including an outer planetesimal disk produced good Solar System analogs, matching most of the criteria derived in Nesvorny (2011).  Therefore, a scenario in which Jupiter "tacked" at $\sim 1.5$ AU when Saturn caught up and was trapped in a 2:1 resonance may also 
explain the evolution of both the inner and outer Solar System.

  Outward migration in 2:1 resonance is possible in disk models that have a very small aspect ratio, typically 
 $h\sim 0.02-0.03$ at the location of Jupiter. For such a value of $h$,  the disk is substantially depleted at Saturn's orbit, creating a strong positive torque exerted on Jupiter and favoring outward migration. This result was confirmed by locally isothermal runs which resulted in a similar outcome for $h=0.02$ and $\alpha=10^{-4}$ .    We note that more 
 realistic protoplanetary disks  models subject to 
 stellar irradiation are thought to have similar aspect ratios at late stages, especially   if their 
 metallicity is small (Bitsch et al. 2014). This suggests  that a Grand Tack scenario with Jupiter and Saturn 
 in 2:1 resonance is possible in evolved protoplanetary disks, but this needs to be checked with 3D 
 radiative hydrodynamical simulations. 

\acknowledgements
Computer time for this study was provided by HPC resources of Cines under allocation c2014046957 made 
by GENCI (Grand Equipement National de Calcul Intensif). We thank the Agence Nationale pour la 
Recherche under grant ANR-13-BS05-0003-002 (project MOJO).

\clearpage



\clearpage









\clearpage

\end{document}